\documentclass[runningheads]{llncs}

\usepackage[T1]{fontenc}
 
\usepackage[utf8]{inputenc}
\usepackage{times}
 
\usepackage{graphicx}
\usepackage{tabularx}
\usepackage{xcolor}
\usepackage{mathrsfs}
\usepackage{dirtytalk}
\usepackage{hyperref}
\graphicspath{ {pics/} }
 
\usepackage{amsmath}
\usepackage{multirow}
\usepackage[symbol]{footmisc}
\usepackage[linesnumbered,ruled]{algorithm2e}

\newcommand*\samethanks[1][\value{footnote}]{\footnotemark[#1]}

\begin{document}

\title{DSS: Synthesizing long Digital Ink using Data augmentation, Style encoding and Split generation}

\author{Aleksandr Timofeev\inst{1}\thanks{work done as a student researcher at Google Research, Z\"urich, Switzerland}\thanks{These authors contributed equally to this work and share first authorship} 
\and Anastasiia Fadeeva\inst{2}\samethanks[2] 
\and Andrei Afonin\inst{1}\samethanks[1]
\and Claudiu Musat\inst{2}
\and Andrii Maksai\inst{2}
}
\institute{EPFL, Lausanne, Switzerland \and
Google Research, Z\"urich, Switzerland}

\maketitle

\begin{abstract}
As text generative models can give increasingly long answers, we tackle the problem of synthesizing long text in digital ink. We show that the commonly used models for this task fail to generalize to long-form data and how this problem can be solved by augmenting the training data, changing the model architecture and the inference procedure. These methods use contrastive learning technique and are tailored specifically for the handwriting domain. They can be applied to any encoder-decoder model that works with digital ink. We demonstrate that our method reduces the character error rate on long-form English data by half compared to baseline RNN and by 16\% compared to the previous approach that aims at addressing the same problem. We show that all three parts of the method improve recognizability of generated inks. In addition, we evaluate synthesized data in a human study and find that people perceive most of generated data as real.
\end{abstract}
\section{Introduction}

With the growing usage of tablets and styluses, handwriting is an increasingly used human computer interaction (HCI) method. 
Recent developments in natural language processing make the interaction increasingly bidirectional. In the past handwriting was mostly used to facilitate the input of information for the human user, making its recognition the primary digital ink task \cite{carbune2020recognition,krishnan2019hwnet}. With the advent of highly capable digital assistants \cite{LaMDA} however, the interactions are becoming ever more natural. One way to add to this immersive HCI experience is to have the digital assistant respond in the same modality as the user input --- handwriting.

Handwriting synthesis is the process of converting printed text labels into handwriting. Traditionally, this was proposed for user-facing features like autocompletion and error correction \cite{Aksan:2018:DeepWriting}. These tasks all need to deal only with short textual sequences --- operating mostly on a word level. For instance, in the case of error correction, only the misspelled word needs to be replaced. Whereas, responses from recent text generative models can contain multiple paragraphs of text. We propose methods to synthesize long inks to accommodate those responses.

The primary trait of synthesized ink is that it needs to be readable in order to be useful to a user. This observation leads to character error rate being the evaluation of choice in the past work on synthesis \cite{Inkorrect,unsupervised}. For applications like autocompletion, similarity between the writer's style and generated ink is critical \cite{Inkorrect,Aksan:2018:DeepWriting}. In the current work we adopt recognisability metric (CER) but the evaluation of stylistic similarity is beyond the scope of this paper.

Some concerns about the synthesis quality are directly linked to the length of the ink. 
Ensuring the stability of the generated ink style is one of them, as it is needed to construct an immersive experience, where the generated ink is consistent and looks real to people. The presence of artifacts is another, as the number of artifacts is linked to the ink size.
We tackle evaluations in both a qualitative and quantitative way, to see which methods of generating long inks are the best and whether they are good enough by human standards.

%What we propose
We propose three distinct methods of improving long generation. To extend the ink generation to long inks we propose data augmentation to bridge the gap between training and test conditions. To improve style consistency for long ink generation we utilize contrastive learning for style transfer. To ensure synthesis generalization to any input length we propose split generation. We show their impact using two different synthesis model architectures --- LSTM and Transformer-based, on two ink representations --- points and Bézier curves. 

We compare ink synthesizer using the three improvements with both internal and external baselines and we observe recognizability improvements ranging from 16\% to 50\%, depending on the architecture. To sum up, 
\begin{itemize}
    \item we create a system that successfully synthesizes long digital ink \footnote{A notebook to test model inference is available here: \url{https://colab.research.google.com/drive/1SB\_vyDcsdSq1CtE9IOD9opBR9IDgG0ly}}
    \item we blend three different approaches: Data augmentation, Style encoder and Split generation (we refer to it as \textbf{DSS}), resulting in large improvements in ink recognizability across multiple ink representations and architectures
    \item we perform an ablation study to quantify the additive individual impact of the three proposed components
    \item we run a user study to strengthen the quantitative analysis with a qualitative one, showing that most of the synthetic ink is perceived as real
\end{itemize}

\section{Related work}
\label{sec:related}
Handwriting synthesis has been a topic of interest for many years \cite{survey}. Digital ink can be represented in images which carry information about stroke style, color, and width, as well as in the background they are drawn on \cite{Alonso2019AdversarialGO,Gan2021HiGANHI} or sequence of coordinates with time information \cite{Graves2013GeneratingSW}. Both of those representations have their benefits. In this study we will focus on sequence-based approaches.

Digital ink synthesis is a generation task where a model receives text as input and outputs a sequence of coordinates that represents handwriting. Methods in this field have evolved from parametric models with a set of handmade style features \cite{LIN20072097} and sigma lognormal modeling \cite{Djioua2008AnIS} to deep neural networks such as LSTM \cite{Graves2013GeneratingSW} and Transformers \cite{NIPS2017_3f5ee243}. Sequence to sequence models are frequently used for machine translation \cite{10.5555/2969033.2969173}, speech recognition \cite{Huber2021InstantOW}, speech synthesis \cite{8462020} and abstractive text summarization \cite{nallapati-etal-2016-abstractive}. These models can be applied to ink generation as well \cite{Graves2013GeneratingSW}. Mixture density network \cite{Bishop1994MixtureDN} are commonly used as the last layer to generate coordinates or Bézier curves \cite{https://doi.org/10.48550/arxiv.2202.12362} to capture a variety of possible next strokes.

Generation of personalized handwriting is of particular research interest due to many appealing applications like spelling correction \cite{Aksan:2018:DeepWriting} and completion \cite{https://doi.org/10.48550/arxiv.2002.10381}. In this setting a model has two inputs – text to generate and style ink. It outputs an ink with the text in the given style. Style determines the appearance of the output ink but doesn't change the content that is written. In the existing approaches, style is usually represented by a vector \cite{Aksan:2018:DeepWriting} or sequence of embeddings \cite{unsupervised}. In this work we adopt style transfer to guarantee style consistency of long digital inks.

Previous studies in other domains suggest that deep neural networks tend to degrade in quality on long-form inputs \cite{anil2022exploring,seq_len}. For recurrent models LSTM \cite{LSTM} block and attention mechanisms \cite{NIPS2017_3f5ee243} were proposed to facilitate long sequence modeling. However, even with those methods the gap in quality between training and out-of-distribution lengths is still present for 
encoder-decoder RNN models \cite{koehn-knowles-2017-six,out_of_dom}.

For transformer models incorporating positional information together with long data in training is a key to modelling long sequences \cite{https://doi.org/10.48550/arxiv.1910.10683,pos,neishi-yoshinaga-2019-relation}. In tasks where long-form training data is unavailable, self-supervised pre-training can help achieve state of the art performance across all lengths \cite{https://doi.org/10.48550/arxiv.1901.02860}. However, this pre-training requires a vast amount of unlabeled data that is not available in the handwriting domain.

While the previous methods focused on improvements in various architectures, they all rely on the existence of these long sequences in the training data.  For the tasks where long training data is unavailable, data augmentation is one way to improve quality for long sequences \cite{9003913,seq_len}. In the case of speech recognition, simulating long-form data improved the quality by 27\% \cite{9003913} and for machine translation data augmentation boosted the performance by 30\% \cite{concat_mt}.

In this paper we will evaluate common architectures for digital ink synthesis on long-form data. We will also investigate the affect of possible improvements like data augmentation and style conditioning on the out-of-domain length quality.

\section{Method}
In this section we propose the methods that focus on improving generalization of encoder-decoder to long sequences models in the digital ink domain. It is especially important because 
handwriting data collection is complex as it needs special equipment such as electronic whiteboard \cite{1575685} or a tablet \cite{Aksan:2018:DeepWriting}. Thus, methods that address the issue without additional data collection are of particular interest.

\subsection{Data augmentation}

\begin{figure}
    \centering
    \includegraphics[width=270px]{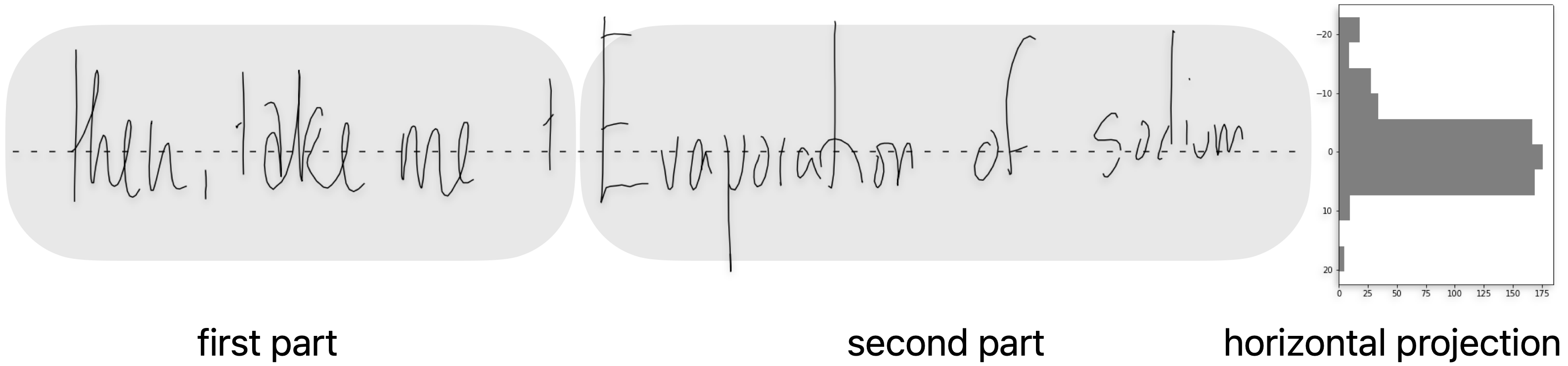}
    \caption{Example of two inks concatenation. In order to put inks on one line, we match the medians of their histograms and scale the second ink to have the same height.}
    \label{fig:concat}
\end{figure}

One of possible approaches to improve performance on longer sequences is to create synthetic long training examples. This idea was explored for speech recognition where speech fragments from the same person were concatenated together \cite{Lu2021InputLM}. Our approach also concatenates inks but instead of samples from the same writer we use stylistically similar inks. Algorithm~\ref{alg:sim} shows our general method of long ink construction from the original training dataset $D$. Thanks to uniform sampling in the algorithm~\ref{alg:sim} (denoted as $U$) we can get a set of long inks by simply calling the function multiple times.
Algorithm~\ref{alg:sim} receives the following inputs:
\begin{itemize}
    \item Style model $f$: receives two inks and outputs a score from $0$ to $1$ which represents the similarity between the writing styles.
    \item Concatenation function: receives two inks and outputs one ink which consists of two parts (example in Fig.~\ref{fig:concat}).
    \item batch size $bs$ controls the diversity of the result dataset, with higher values the probability of overlap between two long inks increases.
    \item Similarity threshold $t$ controls the style consistency, with lower values the chance to concatenate inks with different styles increases.
\end{itemize}
In the following two subsections we describe the style model and concatenation function in more detail.
\begin{algorithm}
    \SetKwInOut{Input}{Input}
    \SetKwInOut{Output}{Output}

    \Input{training dataset $D$, target length $l$, batch size $bs$, threshold $t$, style model $f$, concat function}
    \Output{long ink $r$}
    $r \sim U(D)$\;
    
    \While{$|r|< l$}
    {
        $\text{ink}_{0}$, \dots, $\text{ink}_{bs}$ $\sim U(D)$\;
        $\forall ~ ink \in \text{inks}$: $\text{similarity}(ink) = f(r, ink)$\;

        candidates = $\{i \in 
        \text{inks} ~|~ \text{similarities}(i) \ge t\ or ~ i = \arg\max\text{similarity}\}$\;
        ink $\sim U(\text{candidates})$\;
        $r$=concat($r$,~ink)\;
    }
    return $r$\;
    \caption{algorithm for generating a long sample}\label{alg:sim}
\end{algorithm}

\vspace{-1.0cm}
\vspace{-1.0cm}
\subsubsection{Style model}
We train a model to have similar representations of different texts written in the same style. We use the fact that most inks are stylistically consistent, so we can use different parts of one ink as examples with different text but similar style. To achieve this we use contrastive learning technique. We train a classifier to distinguish between pieces of one ink versus pieces from different inks. We adapt the SimCLR approach from \cite{simclr} where two inputs $x_i$ and $x_j$ are mapped with the same function $g$ into two representations $h_i$ and $h_j$ of a fixed size and then the similarity of these vectors is computed:
$$\text{similarity}(i, j) =  \text{cos}(h_i, h_j)$$
This similarity represents how close in style inks $x_i$ and $x_j$ are. To generate training data for this task we randomly split existing inks into three pieces – beginning part, random gap and final part (we randomly choose two cut points in an ink). 
The random gap is used to generate multiple training examples from one ink. During training we want a similarity of parts from one ink to be close to $1$ and for parts from different inks – to be around $0$. During inference we use a full ink to compute the style representation and subsequently in similarity computation.

\subsubsection{Ink concatenation}
In order to obtain long samples we need a method to merge two inks into one line. In case of misalignment, a training set may lose style consistency, which can lead to model performance degradation. Concatenation can be achieved using a \textit{baseline} – imaginary line on which handwriting is written. We estimate baselines for both inks and apply an rigid transformation to one of them. There are many language specific baseline estimation methods \cite{1030956}, but for the purposes of this study we use a more general method of horizontal projection, shown in Fig.~\ref{fig:concat}. During concatenation we add spaces in between inks which can be adjusted for the languages without space separation like Chinese. Hereby, we generate synthetic long samples from training examples with a similar style.

\subsection{Generation with style conditioning}

Autoregressive models are known to suffer from accumulation of mistakes during inference
\cite{Bengio2015ScheduledSF,prof_forcing}. This problem becomes even more pressing for generation of long sequences. In order to overcome this, we explicitly add a style conditioning into the ink generation model. Conditioning on style is commonly used for style transfer \cite{Aksan:2018:DeepWriting,unsupervised,8683561}, but we use it as a tool to mitigate past mistakes in generation.

\begin{figure}
    \centering
    \includegraphics[width=\textwidth]{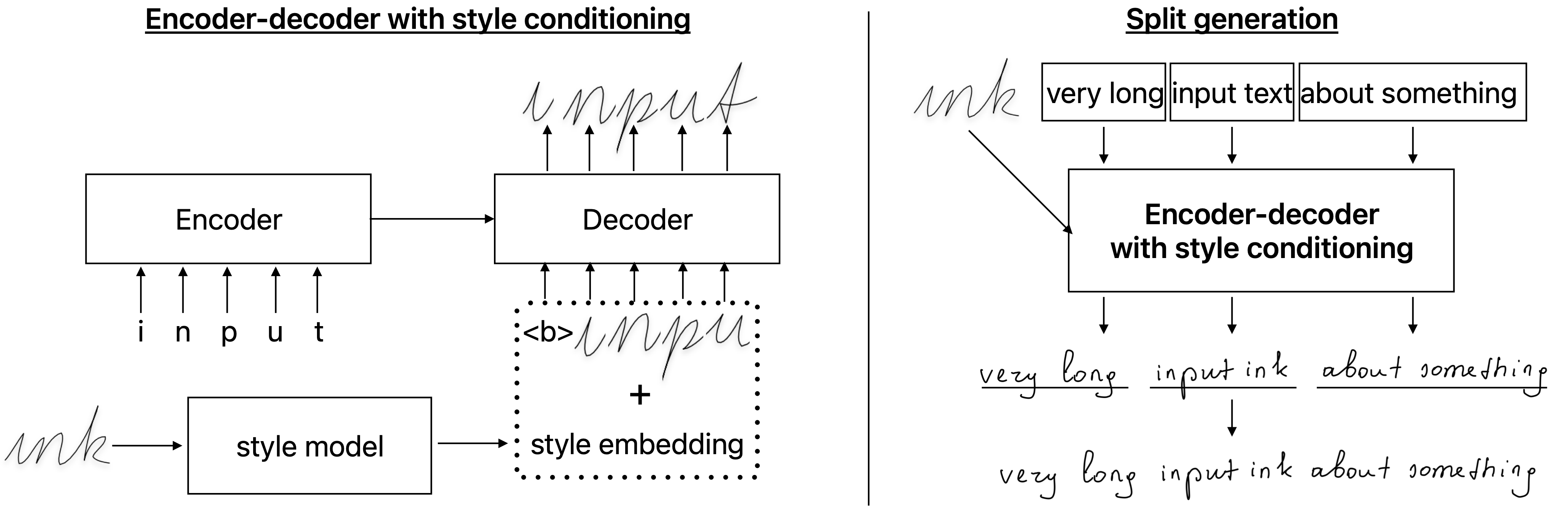}
    \caption{On the left – training of generative model with style conditioning. On the right – split generation procedure with two words.}
    \label{fig:style}
\end{figure}

We propose to train an encoder-decoder model with style conditioning end-to-end, as shown in Fig.~\ref{fig:style}. Encoder inputs the text label and the style model receives an example of ink to generate a style representation vector which is used in the decoding process. Training style model weights at the same time as encoder-decoder parameters helps to adapt style information to a given encoder-decoder model. For model optimisation we combine the negative log likelihood which is typically used to train generative models with SimCLR loss \cite{simclr}:
$$\mathscr{L}(x) = \mathscr{L}_{\text{NLL}} + \mathscr{L}_{\text{SimCLR}} = - \sum_{i \leq n} \log \text{P}(x_i | x_{<i}) + \mathscr{L}_{\text{SimCLR}} (x)$$
The first part of this loss is a likelihood to correctly generate an ink $x$. This part depends on encoder, decoder and style function $f$ weights, as we use style embedding in the decoder's input.
The second part of this loss ensures that function $f$ outputs style representation and doesn't contain information about written text. Similar to the previous section we use the SimCLR approach from \cite{simclr} where two inputs $x_i$ and $x_j$ are mapped with the same function $f$ into two representations $h_i$ and $h_j$ of a fixed size and then the similarity of these vectors is computed as $\text{cos}(h_i, h_j)$. In $\mathscr{L}_{\text{SimCLR}}$ we generate from ink $x$ two pieces $x_{:k}$ and $x_{l:}$ (for some random $k$ and $l$, $k < l$). These pieces are supposed to have high similarity, whereas similarity between $x_{:k}$ and the ink pieces from other inks should be low.

During training we use a piece of target ink as a style input, which can cause target leaking as $\mathscr{L}_{\text{NLL}}$ uses the full target including the style input. However, the use of SimCLR loss promotes independence of style and the text content, therefore no additional steps are needed to prevent target leaking.

During inference we use a separate set of inks for style extraction. Additionally, we use a full ink in the style model rather than just a piece and provide this embedding on each step of decoding as a reminder to the model. This helps avoid the style drift, where the model slowly changes the style of the writing until it becomes completely unrecognizable. Thus, we incorporated a style model into the long ink synthesizer to improve style consistency of long generation.

\subsection{Split generation}
We propose an inference technique for long sequences inspired by dynamic overlapping inference in recognition \cite{out_of_dom}. The idea is to split a long input into separate pieces and calculate their results separately. We split a label at the word boundaries and the maximum number of words in one piece is a hyperparameter of this method, shown in Fig.~\ref{fig:style}. This way we retain the capability to generate cursive handwriting. The main idea of this method is to find the input length that is optimal for model's performance and use it in the inference. However, the main drawback of generating ink pieces separately is the fact that the result parts don't share any information. An example is shown in Fig.~\ref{fig:arrange} random arrangement.

In case of a style mismatch between pieces, people can easily recognize that the ink is synthetic. We propose to use the model with style conditioning from the previous subsection in order to ensure style consistency of the individual generated pieces, that we combine together to obtain the result ink.  To merge pieces together, we match the medians of their horizontal projections, as in the data augmentation step Fig.~\ref{fig:concat}. We don't need to match their scale as they are already similar due to style conditioning.

\begin{figure}
    \centering
    \includegraphics[width=240px]{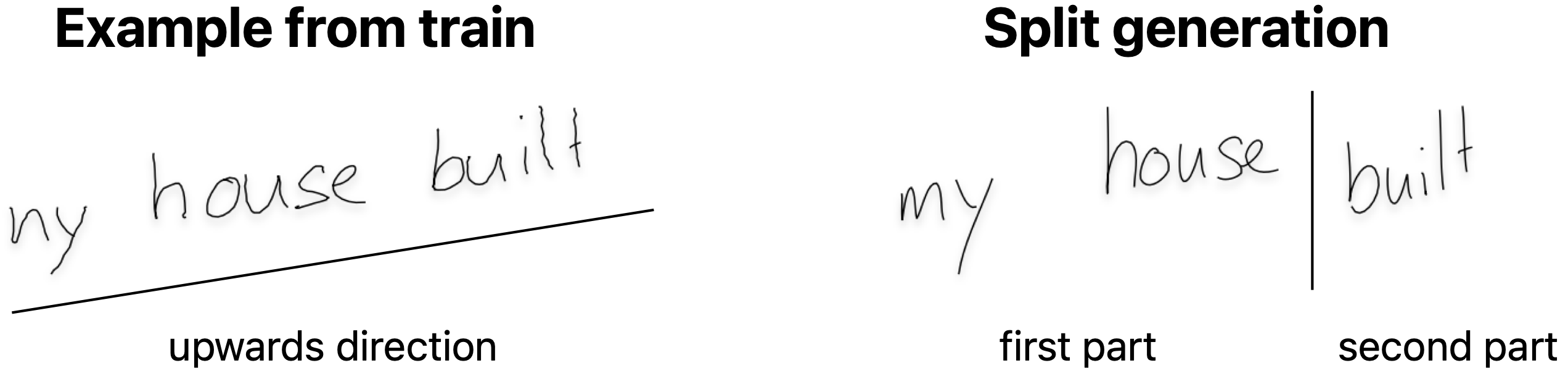}
    \caption{Reason to generate multiple samples and choose the most straight ink in each part of Split Generation.}
    \label{fig:example}
\end{figure}

Mismatch in slope between pieces is also easily recognizable by humans (see Fig.~\ref{fig:example}). To fix this problem we can sample multiple candidates and choose the most horizontal among them using linear regression on the $y$ trajectory. This problem occurs because training inks are not always written horizontally and similar model behavior in split generation leads to mismatched slopes seen in Fig.~\ref{fig:example}.

\subsection{Combining proposed methods}
The proposed methods are complementary to each other as they focus on different parts of an encoder-decoder model. In the training phase, we propose the following order of actions:
\begin{enumerate}
    \item train  a style model with SimCLR loss on dataset $D$
    \item generate an augmented dataset $\hat{D}$ with longer inks from $D$
    \item train an encoder-decoder model with style conditioning on $D \cup \hat{D}$
\end{enumerate}
The final result is the trained encoder-decoder model with style conditioning. In the inference, we suggest the following sequence of steps:
\begin{enumerate}
    \item choose an ink for style extraction
    \item split input text by $n$ words
    \item run inference of the encoder-decoder model on each piece with the same style input
    \item combine pieces into the result ink
\end{enumerate}

\section{Experiments}
\label{sec:results}

\subsection{Setup}
In this section we will apply our modifications to two architectures – \textbf{RNN} \cite{Graves2013GeneratingSW} and \textbf{Transformer} \cite{NIPS2017_3f5ee243}. A classic approach to ink synthesis \cite{Graves2013GeneratingSW} proposed a multi-layer LSTM model with monotonic attention over the labels and gaussian mixture model output for ink synthesis. In our RNN implementation the label encoder is simply one-hot encoding of input characters with dictionary size 70. The decoder consists of an LSTM with 128 units, GMM  monotonic attention layer (size 128, one attention head and 10 components), two LSTM blocks with 256 units, a dense layer and an output GMM model with 10 components \cite{Graves2013GeneratingSW}.

In the transformer model we use a standard encoder-decoder architecture with an output GMM head similar to the RNN model. In both encoder and decoder we use sinusoidal positional embeddings \cite{NIPS2017_3f5ee243}. The transformer label encoder has 4 self-attention blocks (8 heads, 16 units per head, ReLU activation and drop-out rate 0.1). The decoder has 6 standard cross-attention blocks (4 heads, 32 units per head, ReLU activation and drop-out rate 0.1) followed by two dense layers with ReLU activation.

We train both models with batch size of 128 and the Adam optimizer. For the RNN model we use learning rate 1e-4 and for transformer learning rate is 1e-3 with the schedule from \cite{NIPS2017_3f5ee243}. We train each of these models with two different ink representations – raw \cite{Graves2013GeneratingSW} and curve \cite{BSketch}. At inference time, all transformers and the raw RNN models use random sampling of the GMM with bias $\infty$ similar to \cite{Graves2013GeneratingSW}. For the curve RNN we use greedy sampling due to better validation performance. 

\subsubsection{Style encoder}
For style encoding we use a model that consists of one bidirectional LSTM layer with 256 units, followed by a unidirectional LSTM layer with 256 units that outputs the last state and three dense layers on top with sizes 256, 256 and 16. In order to condition a model on a given style we sum a style vector with the decoder input.

\subsubsection{Datasets} We train our models on the \textbf{DeepWriting} dataset which contains more than 34,000 handwritten samples in English \cite{Aksan:2018:DeepWriting}. Examples in the training dataset on average consist of 12 characters and 2.8 words. To evaluate the quality of long generation quantitatively and qualitatively we need a suitable source of handwriting in English. For this purpose, we use \textbf{IAMonDo} which consists of 1,000 pages with English  handwriting, diagrams and drawings in them \cite{Indermhle2010IAMonDodatabaseAO}. We extracted handwriting from those pages and split them into three categories – \textit{long} – more than 7 words, \textit{medium} – 4-7 words and \textit{short} – less than 4 words. For short samples we chose a threshold to match the mean length in the training set (ss shown in Table~\ref{table:datasets}). We then split the rest of the inks into two sets to see progression of quality from in-distribution to increasingly out-of-distribution lengths. For clarity, we include the label sets with our submission \footnote{A notebook with test sets and model inference \url{https://colab.research.google.com/drive/1SB\_vyDcsdSq1CtE9IOD9opBR9IDgG0ly}}.

\begin{table}
\caption{Dataset statistics for train, validation and test datasets.}
\label{table:datasets}
\small
\centering
\begin{tabular}{ |l|l|c|c|c|c| } 
 \hline
 dataset &  & words & mean characters (std) & samples & characters \\
 \hline
 \multirow{2}{*}{DeepWriting} & train & 2.8 (1.2) & 12.41 (5.08)  & 34K & 421K \\
 & valid & 2.8 (1.2) & 12.30 (4.97) & 680 & 8K \\
 \hline
 \multirow{3}{*}{IAMonDO} & long & $>$ 7 & 48.22 (9.06)  & 310 & 15K \\
 & medium & 4-7 & 30.52 (7.98) & 864 & 26K \\
 & short & $<$ 4 & 10.91 (5.63) & 811 & 9K \\
 \hline
\end{tabular}
\end{table}

\subsubsection{Evaluation metric}
Following \cite{arxiv.2202.11456} we measure the recognizability of generated inks with Character Error Rate (CER) on a set of test labels. Our recognizer is an RNN model trained on a private dataset described in \cite{carbune2020recognition}. It performs well on all lengths present in the test sets, see the results on the original data in Table~\ref{table:recognizer_original}.

\subsubsection{Baseline model}
We compare our method to the VRNN model \cite{Aksan:2018:DeepWriting} which generates one character at a time and propagates RNN states between characters to control the style. This model easily generates long input text (see Table~\ref{table:main}) as very limited information is shared between the characters. Another result of this procedure is that characters are rarely connected in the final handwriting. Thus, this model struggles to generate cursive writing.

\subsection{Quantitative results}
In this section we apply proposed changes to two architectures with two different ink representations – raw \cite{Graves2013GeneratingSW} and curve \cite{BSketch}. We train each model 3 times in order to compute the variance between runs. In Table~\ref{table:main} we present the mean CER and standard deviation of the models without any changes and with proposed improvements. Comparing baseline models we see that RNN models degrade on long inputs less than transformers, which matches previous results in the text domain \cite{neishi-yoshinaga-2019-relation}. It is expected as positional embeddings for remote positions are unknown to the transformer model, hurting accuracy, while RNN has monotonic attention which helps with longer sequences. However, even in case of RNN models CER gets 2 times worse for curve setup and 5 times for raw between short and long validation datasets. Thus, all four models can benefit from better out-of-domain length generalisation.

\begin{table}
\caption{Recognizability comparison of our method with baseline models and the VRNN model \cite{Aksan:2018:DeepWriting}. Datasets: long > 7 words, medium 4-7 words, short < 4 words.
}
\label{table:main}
\small
\centering
\begin{tabular}{ |l|l|c|c|c|c| } 
 \hline
 model & & CER long & CER medium & CER short & CER avg \\
 \hline
 VRNN \cite{Aksan:2018:DeepWriting} & & 5.94 (0.22) & 5.05 (0.04) & 4.8 (0.1) & 5.26  \\
  \hline
 \multirow{2}{*}{transformer curve} & DSS & \textbf{3.96 (0.23)} & \textbf{4.0 (0.37)} & 5.3 (0.57) & \textbf{4.42}\\ 
 & base & 60.62 (3.05) & 34.91 (3.36) & 10.35 (0.47) & 35.29\\ 
  \hline
 \multirow{2}{*}{transformer raw} & DSS & 12.73 (1.23) & 9.89 (1.61) & 8.83 (2.06) & 10.48 \\
 & base & 64.34 (1.11) & 41.24 (2.15) & 11.4 (1.4) & 38.99\\ 
  \hline
 \multirow{2}{*}{RNN curve} & DSS & 6.6 (0.39) & 6.03 (0.95) & 7.08 (0.95) & 6.57\\ 
 & base & 12.25 (0.78) & 9.27 (0.93) & 5.29 (0.28) & 8.94\\ 
  \hline
 \multirow{2}{*}{RNN raw} & DSS & 4.42 (0.55) & 4.22 (0.51) & 4.56 (0.48) & \textbf{4.4}\\ 
 & base & 21.48 (5.18) & 11.74 (3.18) & \textbf{4.11 (0.43)} & 12.44\\ 
 \hline
\end{tabular}
\end{table}

In this section we provide details on the hyperparameters utilized for our results and the criteria for their selection. For each of four models we added \textit{10K long samples} built from train dataset and picked the optimal target length $l$ based on validation quality presented in Table~\ref{table:len} (in the appendix). In style condition during inference we used random examples from \textit{DeepWriting validation dataset} as a source of style. This ensured that we did not introduce any stylistic bias compared to unconditional generation, as the distribution of validation dataset matches the training dataset. In order to determine the best number of words in the split generation we applied data augmentation method to validation dataset. We matched the target length of this dataset to the long test dataset. Then, we evaluated models with additional data and style conditioning on the said dataset. Results are presented in Table~\ref{table:split_generation} (in the appendix).

In conclusion, our method significantly outperforms the baseline in all four cases, with curve transformer as well as raw RNN showing the best overall quality. This quality is also better than VRNN model \cite{Aksan:2018:DeepWriting} which requires character segmentation in the training data. This information is absent in most open source handwriting datasets \cite{nguyen2018database}. In our results we show that without any additional annotation, superior quality is attainable across all three target lengths.

\subsection{Ablation study}
In this section we remove each of the three changes that we proposed from the combination and measure the quality to show that all three changes contribute to the optimal model performance. In addition, we do a more detailed ablation study on the curve transformer setup, as it provides the best overall quality.

\subsubsection{Impact of data augmentation.}
In order to measure the impact of synthetic long data, we remove it from training. All of the other parameters stay the same and we train a new set of models only on the original data. This experiment is especially important because split generation significantly reduces the input text length during inference. However, data augmentation improves model performance on short samples as well (see Table~\ref{table:len} in the appendix) possibly due to the moderate size of the training set.

\begin{table}
\caption{Effect of data augmentation.}
\label{table:data_augmentation_effect}
\small
\centering
\begin{tabular}{ |l|l|c|c|c|c| } 
 \hline
 models & data & CER long & CER medium & CER short & CER avg\\
  \hline
 \multirow{2}{*}{transformer curve} & yes & 3.96 (0.23) & 4.0 (0.37) & 5.3 (0.57) & 4.42\\
  & no  &  6.61 (0.5) & 6.89 (0.29) & 7.71 (0.95) & 7.07\\
  \hline
 \multirow{2}{*}{transformer raw} & yes & 12.73 (1.23) & 9.89 (1.61) & 8.83 (2.06) & 10.48\\
  & no & 13.35 (0.59) & 11.42 (1.08) & 11.21 (1.23) & 11.99\\ 
  \hline
 \multirow{2}{*}{RNN curve}  & yes & 6.6 (0.39) & 6.03 (0.95) & 7.08 (0.95) & 6.57\\ 
  & no & 6.78 (0.62) & 6.05 (0.64) & 6.85 (0.64) & 6.56\\ 
  \hline
 \multirow{2}{*}{RNN raw} & yes & 4.42 (0.55) & 4.22 (0.51) & 4.56 (0.48) & 4.4\\
  & no & 4.85 (0.6) & 4.66 (0.83) & 4.94 (1.04) & 4.82\\ 
 \hline
\end{tabular}
\end{table}

In Table~\ref{table:data_augmentation_effect} we can see that for transformers augmented data plays a very important role. The decrease in quality is almost 60\% for curve transformer and around 16\% for raw transformer.
As mentioned in the main results, transformer models fail to generalize to long-form sequences due to constant positional embeddings unseen at training time. The restriction is overcome through the use of synthetic long data. However, for raw RNN model the loss is also quite significant – around 10\%, which shows that augmented data is beneficial in most cases.

\begin{figure}
    \centering
    \includegraphics[width=\textwidth]{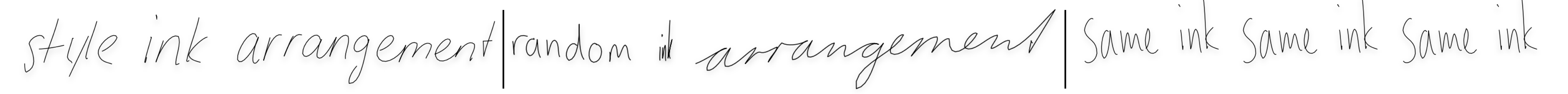}
    \caption{Style, random and same ink arrangements.}
    \label{fig:arrange}
\end{figure}

In DSS we propose to concatenate existing training examples based on their style similarity. We compare this approach to repetition of the same ink many times and concatenation of random samples into one ink. In Fig.\ref{fig:arrange} we can see that random arrangement doesn't preserve style consistency which is an important quality of digital inks. On the other hand, repetition of the same ink returns stylistically consistent inks but lacks diversity in text. These strategies decrease recognizability compared to style-based arrangement in the curve transformer setup, as shown in Table~\ref{table:data_augmentation_diff}.

\begin{table}
\caption{Different data augmentation strategies for curve transformer setup.}
\label{table:data_augmentation_diff}
\small
\centering
\begin{tabular}{ |l|c|c|c|c|c| } 
 \hline
 data arrangement &  CER long & CER medium & CER short & CER avg \\
  \hline
 style & 3.96 (0.23) & 4.0 (0.37) & 5.3 (0.57) & \textbf{4.42}\\
 random & 5.01 (0.52) & 4.96 (0.33) & 5.58 (0.49) & 5.18\\
 repetition & 7.15 (0.34) & 7.02 (0.6) & 8.56 (0.57) & 7.58\\
 \hline
\end{tabular}
\end{table}
\vspace{-1.0cm}

\subsubsection{Impact of style encoding}
\begin{table}
\caption{Effect of style encoding. No style rows show metrics for stylistically inconsistent inks (see random arrangement in Fig.~\ref{fig:arrange}). In priming we prompt a model with the style ink before target ink generation.}
\label{table:no_style}
\small
\centering
\begin{tabular}{ |l|l|c|c|c|c| } 
 \hline
 models & style & CER long & CER medium & CER short & CER avg \\
  \hline
 \multirow{3}{*}{transformer curve} & yes & 3.96 (0.23) & 4.0 (0.37) & 5.3 (0.57) & 4.42\\
  & prime & 6.34 (0.65) & 6.19 (0.7) & 7.39 (1.69) & 6.64\\ 
  & no  & 10.02 (1.77)  & 10.32 (1.04) & 7.73 (0.74) & 9.36\\
  \hline
 \multirow{3}{*}{transformer raw} & yes & 12.73 (1.23) & 9.89 (1.61) & 8.83 (2.06) & 10.48\\ 
  & prime & 17.92 (1.99) & 23.06 (2.31) & 18.74 (2.95) & 19.91\\
  & no  & 13.15 (2.55) & 10.16 (2.56) & 9.81 (2.21) & 11.04\\
 \hline
 \multirow{3}{*}{RNN curve}  & yes & 6.6 (0.39) & 6.03 (0.95) & 7.08 (0.95) & 6.57\\ 
  & prime & 7.49 (0.63) & 6.85 (0.3) & 7.69 (0.69) & 7.34\\ 
  & no & 5.14 (0.95) & 4.45 (0.96) & 4.42 (0.95) & 4.67\\ 
  \hline
 \multirow{3}{*}{RNN raw} & yes & 4.42 (0.55) & 4.22 (0.51) & 4.56 (0.48) & 4.4\\
  & prime & 4.49 (1.46) & 4.25 (1.28) & 4.08 (1.13) & 4.27\\ 
  & no & 3.66 (0.2) & 3.59 (0.86) & 2.16 (0.45) & 3.14\\ 
 \hline
\end{tabular}
\end{table}

Next, we remove style encoder from encoder-decoder architecture, leaving everything else unchanged. In Table \ref{table:no_style} we show results for split generation where we generate label pieces independently and don't share any information between them (rows with no style). In this case split generation returns stylistically inconsistent inks similar to random data augmentation as shown in Fig.~\ref{fig:arrange}.
In a qualitative evaluation, we show that stylistically inconsistent inks are 33\% less likely to be recognized as real. For RNN models character error rate is better without style than with style conditioning, but it is not the case for transformer models where CER gets significantly higher without any style – 110\% for the curves setup.
%79\% of inks generated with style conditioning are consistent in style, but .
%due to the fact that the model is not limited by some specific style.

We also compare these results to a different method of ensuring style consistency, proposed in \cite{Graves2013GeneratingSW}. In split generation we can prime a model on the style ink before generating each piece by teacher-forcing a style input and asking a model to complete it with input text. This way we get result ink with similar style to the style ink. We implicitly determine consistency of the result by using the same style ink in each piece of split generation. The main disadvantage of this approach is that by doing a completion of style ink we increase the target ink length which may result in subpar quality. In Table~\ref{table:no_style} we can see that for curve transformer completion decreases CER by 30\% compared to no style and for RNN raw the quality of completion is similar to style encoding.

\subsubsection{Impact of split generation}
Finally, we evaluate models quality without split generation. Table~\ref{table:no_split} shows that models with fully autoregressive inference perform by 30-100\% worse than with split generation. The gap in quality is especially pronounced for long evaluation sets as we split the labels there into many pieces. Whereas, the quality on short data stays almost the same as in many cases we don't split targets there. However, raw RNN even without split generation still performs similarly to VRNN (see Table~\ref{table:main}).

\begin{table}
\caption{Effect of split generation.}
\label{table:no_split}
\small
\centering
\begin{tabular}{ |l|l|c|c|c|c| } 
 \hline
 model & split & CER long & CER medium & CER short & CER avg\\
  \hline
 \multirow{2}{*}{transformer curve}  & yes & 3.96 (0.23) & 4.0 (0.37) & 5.3 (0.57) & 4.42 \\
  & no  & 10.6 (1.42) & 10.6 (1.42) & 5.33 (0.53) & 8.84\\
  \hline
 \multirow{2}{*}{transformer raw} & yes & 12.73 (1.23) & 9.89 (1.61) & 8.83 (2.06) & 10.48\\ 
  & no & 22.44 (4.37) & 22.4 (2.04) & 9.88 (1.6) & 18.24\\
  \hline
 \multirow{2}{*}{RNN curve} & yes & 6.6 (0.39) & 6.03 (0.95) & 7.08 (0.95) & 6.57\\ 
  & no & 14.82 (2.87) & 10.28 (1.43) & 7.43 (0.97) & 10.84\\ 
  \hline
 \multirow{2}{*}{RNN raw} & yes & 4.42 (0.55) & 4.22 (0.51) & 4.56 (0.48) & 4.4\\
  & no & 7.32 (1.95) & 4.86 (0.95) & 4.54 (0.5) & 5.57\\ 
 \hline
\end{tabular}
\end{table}

In Table \ref{table:best_split} we show that quality can vary quite a bit depending on the number of words in one split. It may seem that generating only one word at a time is the most simple task and that would lead to the best performance. However, we noticed that curve transformer fails to generate one word with punctuation at the end probably due to lack of similar data in training. Thus, it is important to choose number of words that is most suited for each model. This value is consistent for curve transformer between different datasets: long synthetic validation in table \ref{table:split_generation}, long and medium validation in Table~\ref{table:best_split}.

\begin{table}
\caption{Different number of words in split generation for curve transformer setup.}
\label{table:best_split}
\small
\centering
\begin{tabular}{ |l|c|c|c|c|c| } 
 \hline
 n words &  CER long & CER medium & CER short & CER avg\\
  \hline
 1 & 6.3 (0.67) & 5.96 (0.75) & 6.96 (0.97) & 6.41\\
 2 & 4.66 (0.5) & 4.56 (0.33) & 5.62 (0.23) & 4.95\\
 \textbf{3} & 3.96 (0.23) & 4.0 (0.37) & 5.3 (0.57) & \textbf{4.42}\\
 5 & 4.53 (0.01) & 4.83 (0.2) & 5.3 (0.57) & 4.89\\
 \hline
\end{tabular}
\end{table}

%trim=3.5cm 0.5cm 3.5cm 10.5cm,clip,
\subsection{Qualitative evaluation}
\vspace{-0.5cm}
\begin{figure}
    \centering
    \includegraphics[width=\textwidth]{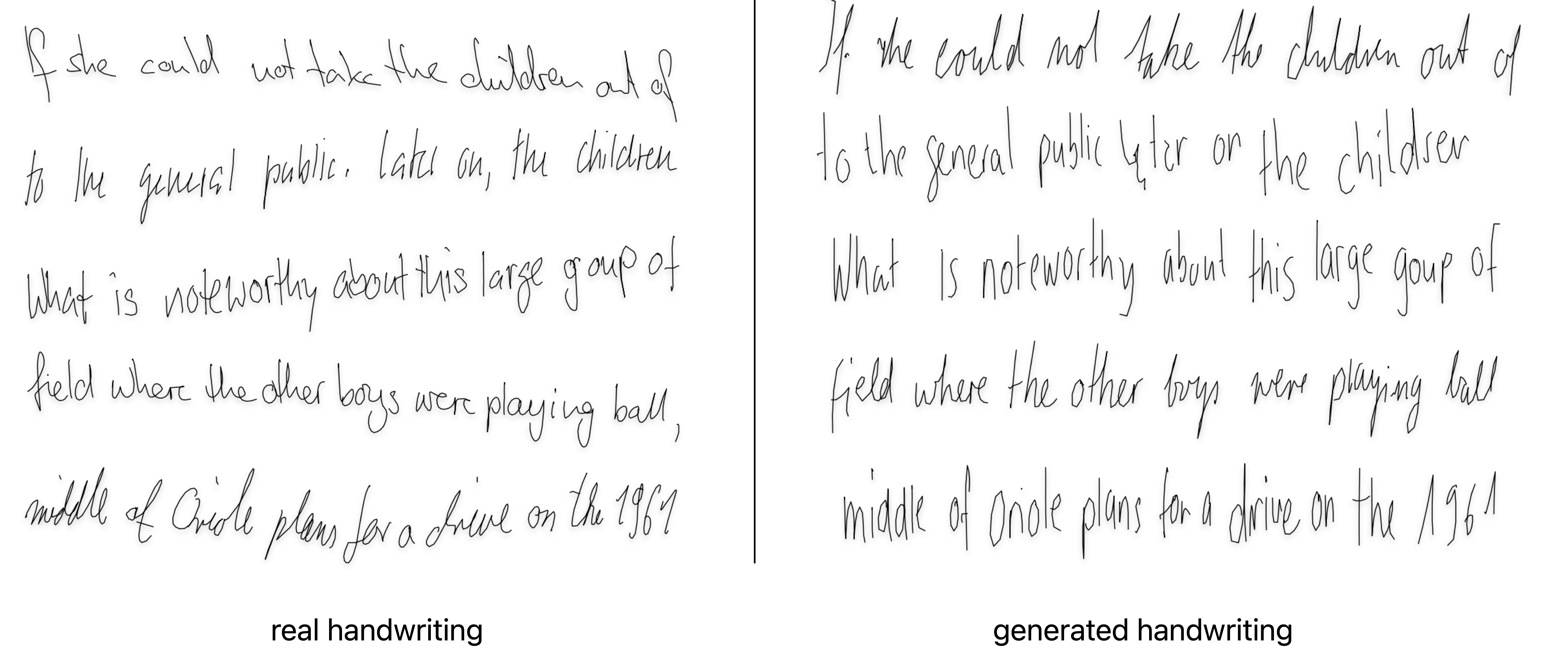}
    \caption{Examples of inks from human study.}
    \label{fig:examples}
\end{figure}
\vspace{-0.5cm}
To ensure that the generated samples are not only recognizable but also look realistic, we performed a  human evaluation. In it, we asked the participants to evaluate a mix of generated and real inks – 100 each with long texts from IAMonDO dataset, see examples in Fig.~\ref{fig:examples}. 
Our model's performance was evaluated on long texts because baseline models struggled the most with generating long samples (see Table~\ref{table:main}). We used a curve transformer model with style encoding, split generation of two words as the direction pattern (see Fig.~\ref{fig:example}) is less pronounced there compared to three words. We also use the sampling of 5 candidates to pick the most horizontal ink. In Table~\ref{table:sampling} in the appendix we show that picking the most horizontal pieces in split generation of two words doesn't significantly decrease recognition quality.

We study whether people can differentiate between real and generated data by showing one ink at a time with a question \say{Does this ink look real?}. We collected 600 responses from 12 people who has worked with digital ink before (3 answers per ink). The results are presented in Table~\ref{table:human_res}. The majority of generated inks were labeled as real – \textbf{79\%}, as a result the F1 score of human answers is only \textbf{0.34}. To check whether original inks are labeled real with higher probability we use Fisher's exact test \cite{fisher_exact_test}. We get the p-value equal to \textbf{0.04}. Thus, we can reject the null hypothesis and conclude that original inks are more likely to be labeled real than generated ones. To sum up, generated inks are frequently perceived as real but there is still a statistically significant difference between the two datasets.
\vspace{-0.5cm}
\begin{table}[h]
\caption{Human study results.}
\label{table:human_res}
\small
\centering
\begin{tabular}{ |l|c|c|c|c| } 
 \hline
 data & look real & consistent style & no artifacts & readable\\
  \hline
 real & \textbf{0.89} & \textbf{0.96} & \textbf{0.77} & \textbf{0.8}\\
 generated & 0.79 & 0.85 & 0.7 & 0.77\\
 \hline
\end{tabular}
\end{table}

\begin{figure}[h!]
    \centering
    \includegraphics[width=\textwidth]{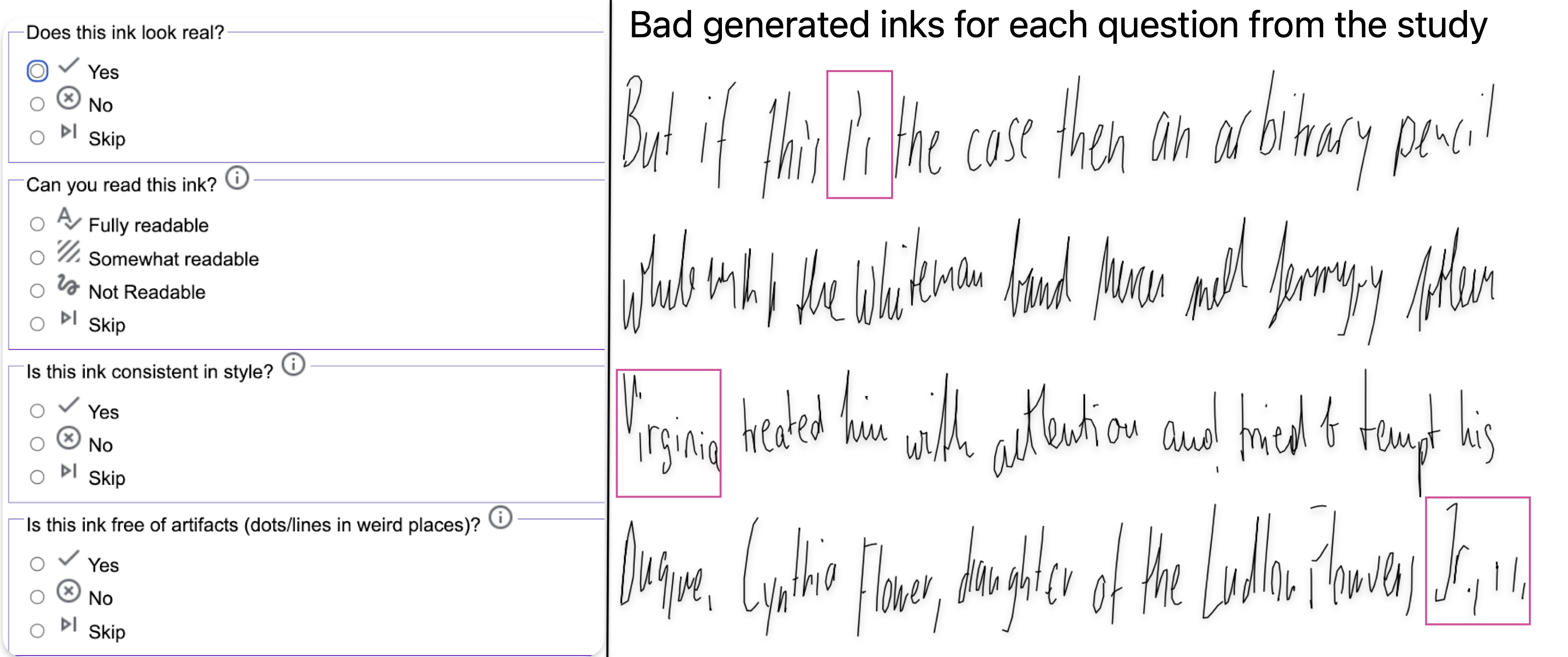}
    \caption{Human study questionnaire on the left and examples of synthesizer's mistakes on the right. The mistakes are marked with rectangles.}
    \label{fig:human_questions}
\end{figure}

Moreover, we asked about the style consistency, presence of artifacts like additional lines or dots and readability of the ink see in Fig.~\ref{fig:human_questions}. In the latter two cases the gap between real and original quality is not as pronounced as in the first two and is not statistically significant. We see that according to people real and generated inks have similar readability which matches our quantitative results (see tables \ref{table:recognizer_original} and \ref{table:sampling}). It is also important to note that generated inks with artifacts are 42\% less likely to be perceived as real and for style inconsistency this number is 33\%.
\section{Conclusion}
\label{sec:conclusion}
We have presented three improvements to encoder-decoder models for digital ink generation – data augmentation, style conditioning and split generation. We have been able to decrease character error rate compared to baseline RNN model by \textbf{50\%} (baseline RNN curve 8.94 vs DSS transformer curve 4.42) and by \textbf{16\%} compared to the previous approach. We've shown that all three proposed methods play an important part in optimal model performance: data augmentation is very important for transformer models, style conditioning has higher recognisability than priming in split generation, split generation has the biggest impact on quality among the three parts of our approach.
In addition, we have conducted a qualitative evaluation where we have verified the quality of generated long inks. People perceive most of synthesized inks as real, but there is still a statistically significant discrepancy between answers in real and generated buckets. We believe that our findings regarding long ink synthesis can be used in real world applications and as a stepping stone for future research in this field.

\pagebreak
\bibliographystyle{splncs04}
\bibliography{references}
\appendix

\section{Appendix}
\vspace{-1.0cm}

\label{sec:appendix}
\begin{table}[ht]
\begin{minipage}[b]{0.5\linewidth}
\centering

\caption{Recognizer CER on test data, caused by data noise and the model's mistakes.}
\label{table:recognizer_original}
\small
\centering
\begin{tabular}{ |c|c|c|c| } 
 \hline
 metric & value\\
 \hline
 CER long & 4.8\\
 CER medium & 4.41\\
 CER short & 5.5\\
 CER avg & 4.9\\
 \hline
\end{tabular}

\end{minipage}\hfill
\begin{minipage}[b]{0.46\linewidth}
\centering
\small
\caption{CER of the curve transformer on long test with and without the most horizontal ink.}
\label{table:sampling}
\small
\centering
\begin{tabular}{ |l|c|c| } 
 \hline
 n words & 1 attempt & 5 attempts \\
  \hline
  1 & 6.29 (0.67) & 7.43 (0.53)\\
  2 & 4.64 (0.48) & 5.0 (0.17)\\
  3 & 3.98 (0.26) & 4.26 (0.32)\\
  5 & 4.53 (0.01) & 4.71 (0.36)\\
 \hline
\end{tabular}
\end{minipage} \hfill
\end{table}

\vspace{-1.0cm}

\begin{table}[ht]
\begin{minipage}[b]{0.5\linewidth}
\centering
\small
\caption{CER for different lengths in data augmentation procedure with candidate pool of 15K and threshold $0.5$. We compare sets with average lengths of 26, 37, 48, 65, 79, 100. For RNN models, training with lengths $> 26$ results in poor performance.}
\label{table:len}
\begin{tabular}{ |p{1.7cm}|c|c| } 
 \hline
 model & mean length & CER valid \\
  \hline
 \multirow{7}{1.8cm}{transformer curves} & - & 8.0 (0.69)\\
  & 26 & 7.46 (1.02)\\
  & 37 & 7.58 (1.56)\\
  & 48 & 6.19 (0.42)\\
  & 65 & 6.98 (0.23)\\
  & \textbf{79} & 6.08 (0.66)\\
  & 100 & 6.62 (0.36)\\
 \hline
 \multirow{4}{1.8cm}{transformer raw }& - & 9.72 (1.85)\\
  & 48 & 9.97 (1.1)\\
  & \textbf{65} & 9.08 (3.68)\\
  & 79 & 11.01 (2.6)\\
 \hline
 \multirow{3}{1.8cm}{RNN curves} & - & 5.94 (0.74)\\
  & \textbf{26} & 5.23 (0.5)\\
 % & 37 & 6.24 (0.77)\\
 \hline
 \multirow{2}{1.8cm}{RNN raw} & - & 3.4 (0.58)\\
 & \textbf{26} & 2.06 (0.64)\\
 \hline
\end{tabular}
\end{minipage}\hfill
\begin{minipage}[b]{0.46\linewidth}
\centering
\small
\caption{CER for different number of words in split qeneration on synthetic long validation dataset.}
\label{table:split_generation}
\begin{tabular}{ |p{1.7cm}|l|c| } 
 \hline
 model & n words & CER long valid \\
  \hline
 \multirow{4}{1.7cm}{transformer curves} & 1 & 8.0 (0.64)\\
  & 2 & 6.03 (0.27)\\
 & \textbf{3} & \textbf{5.57 (0.18)}\\
 & 5 & 5.78 (0.2)\\
 \hline
 \multirow{4}{1.7cm}{transformer raw} & 1 & 17.83 (0.92)\\
 & \textbf{2} & \textbf{14.21 (1.17)}\\
 & 3 & 14.63 (0.23)\\
 & 5 & 20.17 (3.48)\\
 \hline
 \multirow{4}{1.7cm}{RNN curves} & \textbf{1} & \textbf{7.15 (0.87)}\\
 & 2 & 7.47 (0.76)\\
 & 3 & 7.92 (0.95)\\
 & 5 & 9.28 (1.03)\\
 \hline
 \multirow{4}{1.7cm}{RNN raw} & 1 & 7.4 (0.34)\\
 & 2 & 7.17 (0.37)\\
 & \textbf{3} & \textbf{6.66 (0.31)}\\
 & 5 & 7.03 (0.7)\\
 \hline
\end{tabular}
\end{minipage} \hfill
\end{table}

\end{document}